\documentclass[prd,preprintnumbers,floatfix,aps,nofootinbib,notitlepage,showpacs]{revtex4}

\usepackage{graphicx}
\usepackage{latexsym}
\usepackage{epsfig}
\usepackage{amssymb}
\usepackage{epstopdf}

\newcommand{\lp}{\left(}
\newcommand{\rp}{\right)}
\newcommand{\lb}{\left[}
\newcommand{\rb}{\right]}

\newcommand{\ba}{\begin{eqnarray}}
\newcommand{\ea}{\end{eqnarray}}
\newcommand{\be}{\begin{equation}}
\newcommand{\ee}{\end{equation}}

\newcommand{\om}{\omega}

\newcommand{\al}{\alpha}
\newcommand{\bt}{\beta}
\newcommand{\ga}{\gamma}

\newcommand{\la}{\lambda}

\newcommand{\V}{U}

\newcommand{\F}{\Sigma}

\begin{document}

\title{Coupled three-form dark energy}

\author{Tomi S. Koivisto}
\email{tomi.koivisto@fys.uio.no}
\affiliation{Institute of Theoretical Astrophysics, University of Oslo, P.O. Box 1029 Blindern, N-0315 Oslo, Norway}
\author{Nelson J. Nunes}
\email{njnunes@fc.ul.pt}
\affiliation{Faculty of Sciences and Centre for Astronomy and Astrophysics, University of Lisbon, 1749-016 Lisbon, Portugal}
\date{\today}

\begin{abstract}
Cosmology with a three-form field interacting with cold dark matter is considered. In particular, the mass of the dark matter particles is assumed to depend
upon the amplitude of the three-form field invariant. In comparison to coupled scalar field quintessence, the new features include an effective pressure contribution to the field equations that manifests both in the background and perturbation level. The dynamics of the background is analyzed, and new scaling solutions are found. A simple example model leading to a de Sitter expansion without a potential is studied. The Newtonian limit of cosmological perturbations is derived, and it is deduced that the coupling can be very tightly constrained by the large-scale structure data. This is demonstrated with numerical solutions for a model with nontrivial coupling and a quadratic potential.
\end{abstract}


\maketitle

\section{Introduction}

The tiny magnitude of the cosmological constant in the standard $\Lambda$CDM model is not theoretically understood. In attempts to tackle this problem the cosmological constant $\Lambda$ is often replaced by a dynamically evolving entity called dark energy as the agent behind the Universe's present acceleration \cite{Copeland:2006wr}. The theory underpinning the inflationary paradigm that extends the standard big bang model to the earlier Universe also remains to be established. Most often the field responsible for dark energy \cite{Wetterich:1994bg,Amendola:1999er} as well as inflation is a scalar field, but it is of interest to study possible alternatives in terms of other kinds of fields.

Many approaches have been used to reconcile the inherent anisotropies of non-scalar field sources with the isotropy of the cosmological background. Ford's original vector inflation \cite{Ford:1988wq} and some more recent vector dark energy models \cite{Koivisto:2007bp} rely on fine-tuned initial conditions. In the non minimally coupled vector inflation model, it is supposed that there are numerous randomly oriented fields that on the average contribute an isotropic stress energy tensor \cite{Golovnev:2008cf}. The nonminimal couplings of the model have been claimed to introduce pathologies, see 
Refs. \cite{Himmetoglu:2008hx,Himmetoglu:2008zp,Karciauskas:2010as,Golovnev:2011yc} for discussions.
The model has also been generalized to nonminimally coupled 2-form and 3-form inflation \cite{Germani:2009iq,Koivisto:2009sd}. Anisotropies can be avoided by setting a ``triad'' of three vectors pointing in all space-like directions \cite{ArmendarizPicon:2004pm}. By considering more involved models with coupled scalar and vector fields, one may also construct  dynamical systems where the small anisotropy is kept under control \cite{Watanabe:2009ct,Thorsrud:2012mu}. If the vector field is subdominant, the background isotropy can be retained while the field can contribute as a curvaton \cite{Dimopoulos:2011ws} to perturbations in a statistically anisotropic way \cite{Dimopoulos:2008yv}; such models can also be embedded in Type II string theory DBI-type inflation \cite{Dimopoulos:2011pe}. Time-like vector fields are compatible with the FRW symmetries, but a canonical vector is trivial in such a background. Nonminimal couplings and noncanonical kinetic terms have been taken into account in several different contexts  \cite{Eling:2004dk,Koivisto:2008xf,Zlosnik:2007bu,Jimenez:2009ai,Barrow:2012ay} .

The three-form cosmology proposed in 
Refs.~\cite{Koivisto:2009sd,Koivisto:2009ew} can be equivalently described as a noncanonical vector field theory \cite{Gruzinov:2004rq}.  The canonical three-form action can be implemented to successfully model inflation and dark energy \cite{Koivisto:2009fb, Boehmer:2011tp, DeFelice:2012jt} while respecting the FRW symmetry, as the three-form describes a spatial volume element pointing in the orthogonal direction of time. The inflationary models have been found to incorporate large non-Gaussianities in some parameter regions \cite{Mulryne:2012ax} and are claimed to produce more efficient reheating via parametric resonance than usual scalar models \cite{DeFelice:2012wy}. Indeed, very distinct features from scalar models arise when any interactions with other fields are taken into account, including nonminimal couplings to gravity  \cite{Germani:2009iq,Koivisto:2009sd}.  In particular, a new mechanism for magnetogenesis was proposed \cite{Koivisto:2011rm} which is based on the simplest $U(1)$ invariant coupling to the electromagnetic field, and this mechanism turns out to solve the backreaction problem \cite{Demozzi:2009fu} and to be testable as its predictions for non-Gaussianity are extremely sensitive to the model specifics \cite{Urban:2012ib}. Interestingly, such a coupling might be interpreted as an interaction between a hidden sector and the visible photons if the three-form were identified as a hidden sector gauge boson in a framework that is very generic in low energy string theory; moreover such coupling can be further constrained by its rich phenomenology along the lines of e.g., Ref. \cite{Abel:2008ai}.

In the present paper, we investigate couplings of the three-form to the dark matter particles. Interactions between dark energy and dark matter can be justified from high energy physics and phenomenologically they could alleviate the coincidence problem. Coupled scalar field quintessence \cite{Wetterich:1994bg,Amendola:1999er} is very well studied and a myriad of variations has been introduced \cite{Copeland:2006wr}. However, in the three-form context qualitatively new effects will appear, due to the fact that the coupling not only modifies the conservation laws of the interacting components but will also contribute nontrivially to the stress energy tensor, thus modifying the effective source terms for the field equations. Technically this is due to the need to contract the three-form indices in order to render the interaction term covariant. This has been neglected in all previous investigations of vector-field dependent cosmological couplings \cite{Wei:2006tn,Koivisto:2007bp,Zhao:2008tk} and in the three-form models of Ref. \cite{Ngampitipan:2011se}, since in those studies the interactions were not derived from a covariant Lagrangian but introduced as phenomenological terms. Such parameterisations have shortcomings: one cannot uniquely deduce the behavior of perturbations, and they can even result in unphysical instabilities that have not been found to occur in any proper Lagrangian model \cite{Valiviita:2008iv}. What we find here is that in addition to the small-scale instability generic in scalar models\footnote{This instability, first reported and thoroughly analyzed in Ref. \cite{Koivisto:2005nr}, has been dubbed an ``adiabatic 
instability'' \cite{Bean:2007ny} (in the sense of a slowly rolling field), perhaps a bit misleadingly, as it is present in nonadiabatically evolving models as well  \cite{Koivisto:2005nr} and in fact disappears in the exact adiabatic limit \cite{Corasaniti:2008kx,Corasaniti:2010ze}. This linear instability has been considered also in cases where (part of) dark matter consists of neutrinos \cite{Afshordi:2005ym,Mota:2008nj} and was recently shown to exist for a generalized nonconformal form of couplings too \cite{Koivisto:2012za}. It would perhaps be more appropriate to call a Laplacian term like we find here an adiabatic instability (in the usual sense of entropy vs. adiabatic perturbations in cosmology) since, if there were no entropy perturbations, the sound speed squared associated to this term would turn out negative in an accelerating cosmology -- indeed this is what has proven fatal to some unified models of dark matter \cite{Sandvik:2002jz,Koivisto:2004ne}. } the dark matter will also acquire an effective sound speed due to the novel interaction.

The contents of the paper are as follows. In Sec. \ref{fields} we will present the field equations for the system consisting of the three-form field and particles whose mass depends on this field. We model the fluid of particles with the point particle action, which is a sufficient description for most cosmological purposes. In section \ref{backs} we consider these field equations in the FRW space-time, and in particular perform a dynamical system analysis of the phase that generalizes the previous findings. In Sec. \ref{perts} we derive the perturbation equations for the coupled system and analyze it in detail in the Newtonian limit. An evolution equation for the linear structure of matter overdensities is a central result of this study. In Sec. \ref{exams} we consider some specific example models: first, we present a very simple case without potential that provides a new kind of scenario in which to address the dark energy problem, and second, we study a model with a nontrivial coupling and a quadratic potential to demonstrate the large-scale structure implications in this more generic case. Section \ref{concs} briefly concludes the paper. 

\section{Field equations}
\label{fields}

We consider a canonical three-form field $A_{\alpha\beta\gamma}$ with a potential $V(A^2)$ coupled to point particles:
\be \label{lag} 
\mathcal{L} = -\frac{1}{48}F^2 - V(A^2) - \sum_a m_a(A^2)\delta(x-x(\tau))\sqrt{\frac{\dot{x}^2}{g}},
\ee
where $\tau$ is an affine parameter. 
Here $F$ is the field strength tensor $F_{\al\bt\ga\delta}=16\partial_{[\al}A_{\bt\ga\delta]}$.
Thus the mass of the particles depends upon the three-form. 
We restrict here to the conformal form of a coupling, a case that has been thoroughly studied in the context of scalar fields and known to
lead to Yukawa-type corrections to the gravitational potential there; in principle one could construct different coupling forms with the three-form (and a scalar field as well \cite{Koivisto:2012za}), but in this
exploratory study, we will focus on the $m^2(A^2)$ case. 

Because of this interaction, an extra term proportional to the coupling
will appear in the stress energy tensor,
\begin{eqnarray}
 \label{set}
T_{\mu\nu} &=& \frac{1}{6}F_{\mu\al\bt\gamma}F_{\nu}^{\phantom{\nu}\al\bt\gamma} + 6\lb V'(A^2) + 2 \rho_m\lp\ln{m(A^2)}\rp' \rb A_{\mu}^{\phantom{\mu}\al\bt}A_{\nu\al\bt}
-\lb \frac{1}{48}F^2+V(A^2)\rb g_{\mu\nu} + \nonumber \\
&~& \rho_m u_\mu u_\nu\,,
\end{eqnarray}
where a prime means differentiation with respect to $A^2$.
We have assumed that the velocities of the particles are small, so that their energy density $\rho_m$ corresponds to dust,
\be
\rho_m = \sum_a m_a(A^2)\delta^{(4)}(x-x(\tau))\sqrt{\frac{\dot{x}^2}{g}}\,.
\ee
If the extra term
is taken to be included within the matter energy stress tensor, one notes it is no longer a perfect fluid: however, this is a matter of interpretation. The equation of motion for the three-form field follows from the Lagrangian (\ref{lag}) as
\be
\nabla_\al F^{\al\bt\gamma\delta}=12\V'   A^{\bt\gamma\delta}\,,
\ee
where we defined
\be
\V '= V'(A^2) + 2 \rho_m\lp\ln{m(A^2)}\rp'\,.
\ee
Because antisymmetry of the three-form, this further implies that
\be
A^{\alpha\beta\gamma}\V '_{,\al}=-\V'\nabla_\al A^{\al\bt\gamma}\,.
\ee
Using these equations and the vanishing of the divergence of the total stress energy tensor (\ref{set}) then gives us the equation of motion for
the matter fields:
\ba
\nabla_\mu \lp \rho_m u^\mu u_\nu\rp &=& \frac{1}{6} F^{\al\bt\gamma\delta}\lp  \frac{1}{4} \nabla_\nu F_{\al\beta\gamma\delta}-\nabla_\al 
F_{\nu\bt\gamma\delta}\rp - 2\V ' \lp F_{\nu\al\bt\gamma}A^{\al\bt\gamma} + 3 A^{\al\bt\gamma}\nabla_\al A_{\nu\bt\gamma}\rp  + \nonumber \\
&~& 2V' A^{\al\bt\gamma}\nabla_\al A_{\nu\bt\gamma}\,.
\ea
Using the antisymmetry of the three-form and its field strength, this further simplifies to
\be
\nabla_\mu \lp \rho_m u^\mu u_\nu\rp =-4\lp\ln{m(A^2)}\rp' A^{\al\bt\gamma}\nabla_{\nu}A_{\al\bt\gamma}\,.
 \ee
This closes the system of equations. It also seems that the coupling vector $\nabla_\mu {T_{\rm (matter)}}^\mu_\nu$ is proportional neither to the
four velocity of matter $u_\nu$ nor to the four-velocity of the three-form field $\sim e_{\nu\al\bt\gamma}A^{\al\bt\gamma}$. This feature is not captured
by the phenomenological parameterisations considered in the literature, e.g. Refs. \cite{Valiviita:2008iv, Honorez:2010rr}.

%%%%%%%%%%%%%%%%%%%%%%%%%%%%%%
\section{Background cosmology}
\label{backs}

The ansatz 
\be
A_{ijk}=a^3(t)\epsilon_{ijk}X(t)
\ee
is compatible with the FRW symmetries since it corresponds to the time-like component of the dual vector field. We normalize by the cube of the scale factor $a(t)$ so that $A^2=6X^2$. The potential and the coupling can then be regarded as functions of the effective field $X$. In terms 
of that field, we define the shorthand notations for convenience,
\be
f\equiv\frac{2}{\kappa}\left(\ln{m}\right)_{,X}\,, \hspace{1cm} \la \equiv \frac{1}{\kappa}\left(\ln{V}\right)_{,X}\,, 
\ee
where $\kappa^2 = 8 \pi G$.  The Friedmann equations are
\ba
H^2 & = & \frac{\kappa^2}{3}\lb \frac{1}{2}\lp \dot{X}+3HX\rp^2+ V + \rho_m \rb\,, \\
2\dot{H}+3H^2 & = &  \kappa^2 \lb \frac{1}{2}\lp \dot{X}+3HX\rp^2 + V -  \lp V_{,X} + \kappa \rho_m f\rp X \rb\,. \label{pres}
\ea
The equation of motion for the three-form field is
\be \label{k-g}
\ddot{X}=-3H\dot{X}-3\dot{H}X-V_{,X}-\kappa\rho_m f\,,
\ee
and the continuity equation for the matter density is
\be \label{cont}
\dot{\rho}_m+3H\rho_m=\kappa\rho_m f\dot{X}\,.
\ee
As discussed in the introduction, a peculiar feature here is that, in addition to the coupling terms in the conservation equations (\ref{k-g}) and (\ref{cont}), there is an extra pressure contribution in the second Friedmann equation (\ref{pres}). This can also be seen in the form of the equation of state parameter $w_X$,
\begin{equation}
\label{eqofstate}
w_X = -1 + \frac{V_{,X} + \kappa f \rho_m}{\rho_X} X.
\end{equation}

%%%%%%%%%%%%%%%%%%%%%%%%%%%%%%
\subsection{Dynamical system analysis}

We define the dimensionless variables
\be
x=\kappa X\,, \hspace{0.7cm} y= \frac{\kappa}{\sqrt{6}}\lp X'+ 3X\rp\,, \hspace{0.7cm} z^2 = \frac{\kappa^2 V}{3H^2}\,, \hspace{0.7cm} \omega^2 = \frac{\kappa^2 \rho_m}{3H^2}\,. 
\ee
Here and in the following, the prime stands for the derivative with respect to $e$--folding time.
In terms of these variables, the Friedmann constraint reads
\be 
1=y^2+z^2+\omega^2\,.
\ee
This allows us to eliminate one of the variables. We shall consider the phase space spanned by $x$, $y$ and $z$. The autonomous
system of evolution equations can be written as
\ba
x'&=& -3x + \sqrt{6}y\,, \\
y'&=& \frac{1}{2} \left[f \left(\sqrt{6}-3 x y\right) \left(y^2+z^2-1\right)+3 y \left(z^2 (\la x-1)+1\right)-\sqrt{6} \la z^2-3 y^3\right]\,,\\
z' &=&\frac{z}{2}  \left[-3 (f x+1) \left(y^2+z^2-1\right)+3 \la x \left(z^2-1\right)+\sqrt{6} \la y\right]\,.
\ea
There are four types of fixed points. 
\begin{itemize}
\item {\bf Type A}: Matter domination. Strictly this fixed point exists only when $f(0)=0$. It is always unstable since the eigenvalues are
$(\frac{3}{2}, -\frac{1}{4}(3 \pm \sqrt{81 - 48 f_{,X}(0)}))$. The effective equation of state is of course
$w_{\rm eff}=0$ for this solution.
\item {\bf Type B}: Three-form saturated de Sitter. This is the cosmological constant which a massless three-form generates. 
Then $x=\pm \sqrt{\frac{2}{3}}$, $y=1$ and $\omega=z=0$. The eigenvalues are $(-3,-3,0)$, the vanishing eigenvalue corresponding to the perturbations in the direction of $z$. Plugging perturbations only in this direction and expanding to the first nonvanishing order, we get
$\delta z' = -\frac{1}{2}[3 + \sqrt{6}(\lambda(\sqrt{\frac{2}{3}})+f(\sqrt{\frac{2}{3}})) ] \delta z^3$. 
When the expression in the square brackets
is positive, this fixed point is an attractor, otherwise a saddle point. The stability thus depends upon both the functions $f$ and $\lambda$.
\item {\bf Type C}: Potential dominated de Sitter\footnote{Note that this coincides with the type $B$ fixed point when $z=0$. Also, one could
extend the solutions to $x_0^2>\frac{2}{3}$ if negative potentials were allowed, corresponding to imaginary $z$ here.}. This fixed point exists when $\la=0$, i.e. at the extremum of the potential at $x_0$. Then $y=\sqrt{\frac{3}{2}}x_0$ and $z=\sqrt{1-y^2}$. The eigenvalues are
$(-3, -\frac{1}{2}(3 \pm \sqrt{9 - 3\lambda_{,X}(x_0)(2 - 3 x_0^2)^2})$. In accordance with our intuition, the fixed point is stable if the effective field $X$ has a real mass at the minimum, $V_{,XX}(x_0)>0$. 
\item {\bf Type D}: Scaling solutions. The scaling solution for which $\om^2 =\frac{3}{2}|x_0/f|$ exists only in the presence of the coupling for 
$x_0^2<\frac{2}{3}$. The Universe expands as matter dominated $w_{\rm eff}=0$ and $y=\sqrt{\frac{3}{2}}x_0$, $z=0$. For this point to be stable, we need
$f(x_0)x_0<-1$. In addition, $f$ cannot be a constant but
\be
\frac{3\lp 2 - 9 x_0^2 - 4 f(x_0) x_0 (3 x_0^2-2)\rp}{(2 - 3 x_0^2)^2} < f_{,X}(x_0) < 
\frac{3}{16}\lp 9 + f(x_0) x_0 \frac{18 + 9 f(x_0) x_0 + 16}{2 - 3 x_0^2} \rp\,.
\ee
\end{itemize}
This concludes the quite elegant summary and generalization of all previous results.

%%%%%%%%%%%%%%%%%%%%%%%%%%
\section{Cosmological perturbations}
\label{perts}

We parametrize the scalar fluctuations of the three-form using the two scalars $\alpha$ and $\alpha_0$ \footnote{We correct here a typographic error made in Ref.\cite{Koivisto:2009fb} in the definition of $A_{0ij}$.},
\begin{eqnarray}
A_{0ij} &=& a \epsilon_{ijk} \alpha_{,k} , \\
A_{ijk} &=& a^3 \epsilon_{ijk} (X + \alpha_0).
\end{eqnarray}

The equation of motion, $\nabla_\alpha F^{\alpha \beta \gamma\delta} = 12 (V'(A^2)+ 2 \rho (\log m(A^2))' A^{\beta\gamma\delta}$, yields the equation of motion for $\alpha_0$,
\begin{eqnarray} \label{eoma}
\ddot\alpha_0 &+& 3 H \dot\alpha_0 + (3 \dot H + V_{,XX})\alpha_0 -\frac{\nabla^2}{a^2}(\dot\alpha-2 H \alpha)+ (\dot X + 3 H X)(3 \dot \phi -\dot \psi) + \nonumber \\ 
&~& 3 (V_{,XX}X-V_{,X})\phi + 2 V_{,X}\psi = \kappa f \rho_m \left(\delta_m - 3 \phi + 2\psi\right) - \kappa f_{,X}\rho_m(\alpha_0+ 3 X \phi),
\end{eqnarray}
where $\delta_m=\delta\rho_m/\rho_m$, and  the constraint
\begin{eqnarray} \label{eoma2}
\dot\alpha_0 + 3 H \alpha_0 + (\dot X + 3 H X) (3 \phi-\psi) + 
\left(\frac{V_{,X}}{X} - \frac{\nabla^2}{a^2}\right)\alpha  = - \frac{\kappa f \rho_m}{X} \alpha.
\end{eqnarray}

The identity $\nabla\cdot(\nabla \cdot F)=0$ gives the additional constraint
\begin{eqnarray}
\label{antisymcond}
&~& \frac{\partial}{\partial t} \left( \frac{V_{,X}\alpha}{X}\right) - V_{,XX}(\alpha_0+3 X \phi) - V_{,X}\psi = \nonumber \\
&~&  \kappa f \rho_m \left(\delta_m + \psi \right) + \kappa f_{,X} \rho_m (\alpha_0 + 3X\phi) -
 \frac{\partial}{\partial t} \left( \frac{\kappa f \rho_m}{X} \alpha \right)\,.
\end{eqnarray}
For the matter perturbations, we obtain
\begin{eqnarray}
\dot\delta_m + \frac{\theta}{a} - 3\dot\phi = \frac{\partial}{\partial t} \left( \kappa f (\alpha_0 + 3 X \phi) \right),
\end{eqnarray}
and
\begin{eqnarray}
\frac{\dot\theta}{a} + \frac{\nabla^2}{a^2} \psi + H \frac{\theta}{a} = -\kappa f \left( \dot{X} \frac{\theta}{a} + \frac{\nabla^2}{a^2} (\alpha_0 + 3 X\phi) \right).
\end{eqnarray}
By combining these two we obtain
\begin{eqnarray} \label{eomm}
\ddot\delta_m + 2 H \dot \delta_m - 6 H\dot \phi - 3\ddot\phi - \frac{\nabla^2}{a^2} \psi &=& \left( 2 H + \kappa f \dot X + \frac{\partial}{\partial t}\right) \frac{\partial}{\partial t}\left( \kappa f (\alpha_0+ 3X \phi)\right) \nonumber \\
&+& \kappa f \left(\dot{X}(3\dot\phi-\dot\delta_m) + \frac{\nabla^2}{a^2}(\alpha_0+3X\phi)\right).
\end{eqnarray}

For the Einstein tensor we have that 
\begin{eqnarray}
\label{G00}
\frac{1}{2}\delta G^0_0 &=&  3 H (\dot\phi+H\psi)-\frac{\nabla^2}{a^2}\phi , \\
\label{G0i}
\frac{1}{2}\delta G^0_i &=&  - \dot\phi_{,i} - H \psi_{,i}, \\
\label{Gii}
\frac{1}{2}\delta G^i_i &=& \ddot\phi +H (3\dot\phi+\dot\psi)+ (2\dot H + 3 H^2)\psi - \frac{1}{3} \frac{\nabla^2}{a^2}(\phi-\psi) , \\
\label{Gij}
\frac{1}{2}\delta G^i_j &=& \frac{1}{2a^2} (\phi-\psi)_{,ij},
\end{eqnarray}
and the perturbed components of the energy-momentum tensor read
\begin{eqnarray}
\label{T00}
-\delta T^0_0 &=& \delta \rho = \delta\rho_m + (\dot X+ 3 H X)\left(\dot\alpha_0 + 3 H \alpha_0 +(\dot X + 3 H X) (3\phi-\psi)- 
\frac{\nabla^2}{a^2}\al \right) + \nonumber \\
&~& V_{,X}(\alpha_0+3X\phi), \\
\label{T0i}
\delta T^0_i &=& -(\kappa f \rho_m + V_{,X})\alpha_{,i} + a \rho_m v^i, \\
\label{Tii}
\delta T^i_i &=& \delta p = - (\dot X+ 3 H X)\left(\dot\alpha_0 + 3 H \alpha_0 +(\dot X + 3 H X) (3\phi-\psi)- \frac{\nabla^2}{a^2}\al \right) + \nonumber \\ 
&& V_{,XX} X (\alpha_0+3 X\phi)+ 
 \kappa f \rho_m \left(\delta_m + \alpha_0 + 3 X \phi\right) + \kappa f_{,X} X \rho_m (\alpha_0+3 X \phi), \\
\label{Tij}
\delta T^i_j &=& 0.
\end{eqnarray}

Einstein's equation $\delta G_i^j = \kappa^2 \delta T_i^j$ gives
\begin{equation}
\psi = \phi.
\end{equation}
Using this result and equations (\ref{G0i}) and (\ref{T0i}) and eliminating $\dot \alpha_0$ with Eq. (\ref{eoma2}) give, through Einstein's equation $\delta G_i^0 = \kappa^2 \delta T_i^0$,
\begin{eqnarray}
\label{e0i}
-\frac{\nabla^2}{a^2} (\dot\phi+H\psi)= \frac{\kappa^2}{2}\left( \rho_m \frac{\theta}{a}-(\kappa f \rho_m + V_{,X}) \frac{\nabla^2}{a^2} \alpha \right).
\end{eqnarray}
Using Eqs. (\ref{G00}) and (\ref{T00}) and equating  $\delta G_0^0 = \kappa^2 \delta T_0^0$, yield
\begin{eqnarray}
\label{e00}
&& -\frac{\nabla^2}{a^2} \phi+ 3 H(\dot \phi + H \phi) - \frac{\kappa^2}{2}V_{,X} \left[\frac{1}{X}(\dot X + 3 H X) \alpha - \alpha_0 - 3X \phi\right] + \frac{\kappa^2}{2} \rho_m \delta_m 
= \nonumber \\
&& \frac{\kappa^2}{2} \kappa f \rho_m \frac{1}{X} (\dot X + 3 H X) \alpha.
\end{eqnarray}
Finally, Eqn.~(\ref{Gii}) and (\ref{Tii}) through $\delta G_i^i = \kappa^2 \delta T_i^i$ give
\label{eii}
\begin{eqnarray}
\ddot \phi &+& 4 H \dot \phi + (2 \dot H + 3 H^2)\phi - \frac{\kappa^2}{6}(\dot X + 3 H X)\left[ \dot \alpha_0 + 3 H \alpha_0 - \frac{\nabla^2}{a^2}\alpha - 2 (\dot X+ 3H X)\phi\right] - \nonumber 
\\
&-& \frac{\kappa^2}{6} X V_{,XX} (\alpha_0+3 X \phi) = -\frac{\kappa^2}{6} \rho_m (\kappa f + \kappa f_{,X}X) (\alpha_0+3 X \phi) - \frac{\kappa^2}{6} \kappa f X \rho_m  \delta_m. 
\end{eqnarray}
In the limit of vanishing coupling these equations reduce to those in Ref. \cite{Koivisto:2009fb}.

%%%%%%%%%%%%%%%%%%%%%%%%%%%
\subsection{Newtonian limit of perturbations}
Let us then consider Newtonian scales in the late Universe.
Going now to Fourier space and using Eqs. (\ref{e0i}) and (\ref{e00}) in combination, we can write an equation for $k^2 \phi$ in terms of the density contrast in the comoving gauge, $\delta_m^c$ (i.e., when $\theta = 0$),
\begin{equation}
\label{poisson1}
\left(\frac{k}{a}\right)^2 \phi = -\frac{\kappa^2}{2} \left( \rho_m\delta_m^c + V_{,X}(\alpha_0+3 X \phi) - \frac{\dot X}{X} (V_{,X}+\kappa f \rho_m)\alpha \right).
\end{equation}
The comoving
density perturbation $\delta_m^c \approx \delta_m$ coincides with that in the Newtonian gauge in the present approximation.
Constraint (\ref{e0i}) also tells us that $H\alpha \sim X \phi$ which means that in the small scale limit, for $k/a \gg H$, we can neglect the last term of Eq. (\ref{poisson1}). Keeping this in mind we can use the  anti-symmetry condition (\ref{antisymcond}) to obtain a relation for $\alpha_0+3 X \phi$,
\begin{equation}
\alpha_0 + 3 X\phi = - \frac{\kappa f \rho_m}{V_{,XX} + \kappa f_{,X}\rho_m} \delta_m + {\cal O}(X\phi,H\alpha).
\end{equation}
Substituting this expression in Eq. (\ref{poisson1}) and neglecting the terms proportional to $X\phi$ and $H\alpha$ as they are small when compared to the $(k/a)^2\phi$ term for sub-horizon scales, 
we obtain the modified Poisson equation
\begin{equation}
\left(\frac{k}{a}\right)^2 \phi = - \frac{\kappa^2}{2}\rho_m\left( 1 + \frac{V_{,X}\F}{\kappa f}\right)\delta_m,
\end{equation}
where
\begin{equation}
\label{deff}
\F \equiv - \frac{\kappa^2 f^2 \rho_m}{V_{,XX} + \kappa f_{,X} \rho_m}.
\end{equation} 
Similarly, we can substitute for $\kappa f (\alpha_0 + 3 X \phi) \approx \F \delta_m$ in the equation for the matter density contrast (\ref{eomm}) to obtain
\be
\label{deltaeq}
\ddot{\delta}_m+ \left( 2H + \kappa f \dot X - \frac{2 \dot \F}{1-\F}\right)\dot{\delta}_m = 
\frac{\kappa_{\rm eff}^2}{2}\rho_m\delta_m - \frac{k^2}{a^2}c_{\rm eff}^2\delta_m\,. 
\ee
We obtained a modified friction term. In addition, the dark matter perturbation feel an effective Newton's constant given by the 
expression
\be
\kappa_{\rm eff}^2 = \frac{\kappa^2}{1-\F} \left[ 1 + \frac{2}{\kappa^2 \rho_m}\left( \ddot \F + (2 H + \kappa f \dot X)\dot \F - \frac{\kappa^2}{2}\frac{V_{,X}\kappa f \rho_m}{V_{,XX}+\kappa f_{,X}\rho_m}\right)\right].
\ee
Furthermore, the coupling introduces an effective sound speed given by
\be
c_{\rm eff}^2 = \frac{\F}{1-\F}\,.
\ee

Recall that the quantity $\F$, which vanishes when the coupling is zero, was defined in Eq. (\ref{deff}). 
Because of these modifications in the formation of large-scale structure, we expect that the coupling should be quite small in order to produce a viable
matter power spectrum. This has been found to be the case for other cosmological models featuring effective sound speed terms, such as
unified models of dark matter \cite{Sandvik:2002jz,Koivisto:2004ne} and some modified gravity models \cite{Koivisto:2006ie,Li:2007xw}. If the sound speed squared is positive, there will be oscillations at small scales of the matter power spectra, which are stringently constrained by observations. We have positive $c_{\rm eff}^2$ when 
\be
\frac{1}{\kappa}f_{,X} < -\lp f^2 +  \frac{V_{,XX}}{\kappa^2\rho}\rp\,   \hspace{1cm}  \text{i.e.} \hspace{1cm}  f_{,x} < -\lp f^2 +  \frac{V_{,xx}}{\rho}   \rp\,.
\ee
Thus, there is an instability at small scales unless the slope of the coupling is negative enough. When very tiny, such a coupling could be compatible with observed structures and perhaps help to alleviate the possible lack of predicted small-scale structure in the $\Lambda$CDM model. The study of specific models and their observational constraints from structure formation is outside the scope of the present paper. In the following section we only consider the background evolution of the perhaps simplest scenario using a coupling. More general investigation including the computation of CMB and LSS will be carried out elsewhere. 

%%%%%%%%%%%%%%%%%%%%%%%%%%%%%%%
\section{Example models}
\label{exams}

We will look analytically and numerically at two example models, the first specified by a constant coupling $f$ and no potential energy, and the other with a coupling and a quadratic potential.

\subsection{Model with no potential energy} 

\begin{figure}[h]
\begin{center}
\includegraphics[width=0.8\columnwidth]{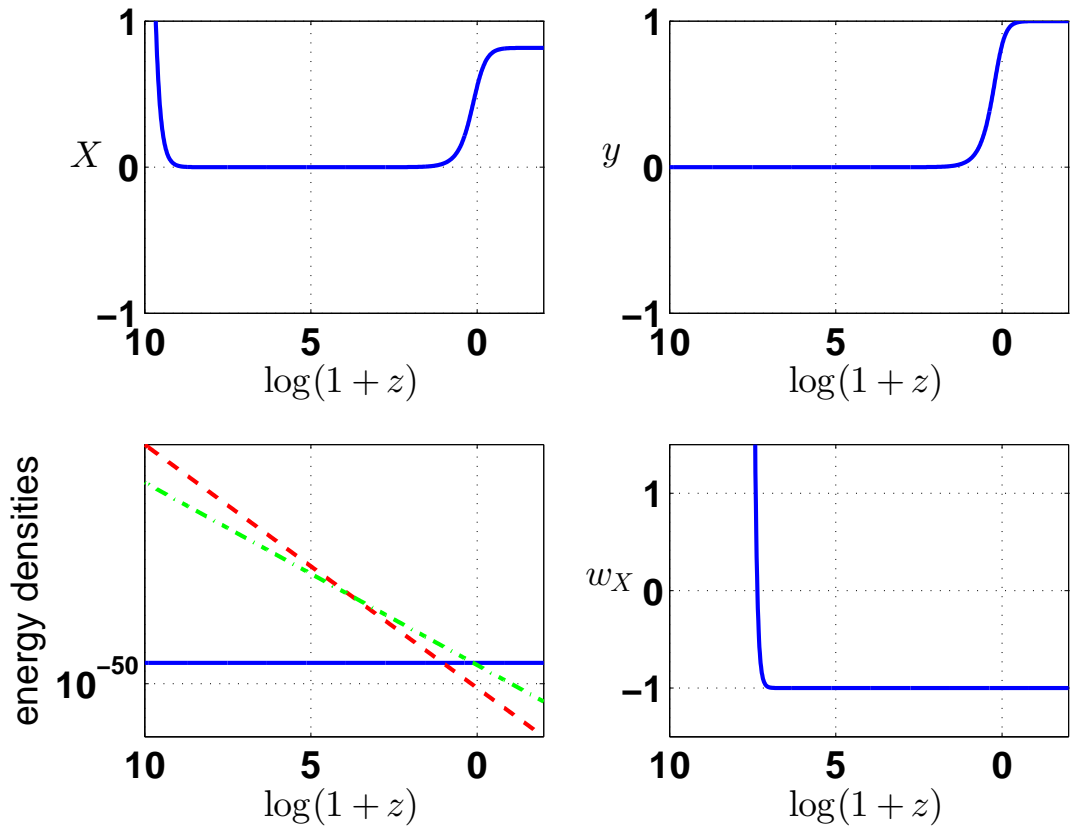}
\caption{\label{fig:case2} Dynamics for a model without potential but with a linear coupling, $f = 1.62\times10^{-12}$, $\kappa X_i = 10$, and $y_i = 1.18 \times 10^{-18}$. The different panels show the evolution of $X$, $y$; the energy density of the various components of the Universe; and the equation of state parameter $w_X$ with redshift. In the lower-left panel, the lines represent the energy density of radiation (dashed line), dust (dotted-dashed line) and $X$ (solid line). In this scenario, the Universe evolves from the fixed point $A$ to the fixed point $B$.}
\end{center}
\end{figure}

Here we will present a very simple realization of a three-form accelerated cosmology that may also address some of the fine-tuning problems of dark energy.
We can accelerate the Universe without a potential. In fact, a massless three-form is nothing but a cosmological constant\footnote{This fact was used by Hawking in an approach to explain the vanishing of the cosmological constant problem \cite{Hawking:1984hk}. The canonic term of the massless field $A$ is the ``gauge fixing" term for the dual vector $\star A$, $F^2(A) \sim (\nabla \cdot \star A)^2$. Maxwellian electromagnetism extended by adding this term to the fundamental electromagnetic Lagrangian provides an electromagnetic origin for the cosmological constant \cite{Jimenez:2008nm}; if this contribution was excited as a quantum fluctuation during inflation the predicted value could match with the observed one \cite{Jimenez:2009sv}. }.   
The de Sitter expansion in the asymptotic future is due to the cosmological constant term
generated by the massless three-form. It automatically adjusts to the current Hubble rate, so as long as we get the transition going on around the 
present epoch, the scale of the effective cosmological constant is naturally of the observed magnitude. We consider a scenario where this is achieved by the coupling. 
The coupling will, however, set the field moving, and eventually it will reach the attractor $B$. 
Let us consider the simple case  $m(A^2) \sim \exp[\kappa f_0 \sqrt{A^2/6}]$ so that $f=f_0$ is a constant\footnote{In this case however, the Newtonian limit derived above might not be valid since when defining  (\ref{deff}) we have assumed that $V_{,XX}+f_{,X}\kappa\rho \neq 0$.}. 
Setting the initial conditions at $1+z_i = 10^{10}$ and fixing 
$f=1.62\times 10^{-12}$ gives us the realistic evolution depicted in Fig.~\ref{fig:case2}. 
In this example,  the field $X$ never crosses zero, and therefore, the equation of state parameter $w_X$ never becomes smaller than $-1$. However,  the initial ratio $\rho_m/\rho_X \ll  f$, which means, by inspection of Eq.~(\ref{eqofstate}), that $w_X$ must be initially much larger  than unity, which is indeed observed in the lower right panel of Fig.~\ref{fig:case2}.

\begin{figure}[h]
\begin{center}
\includegraphics[width=0.8\columnwidth]{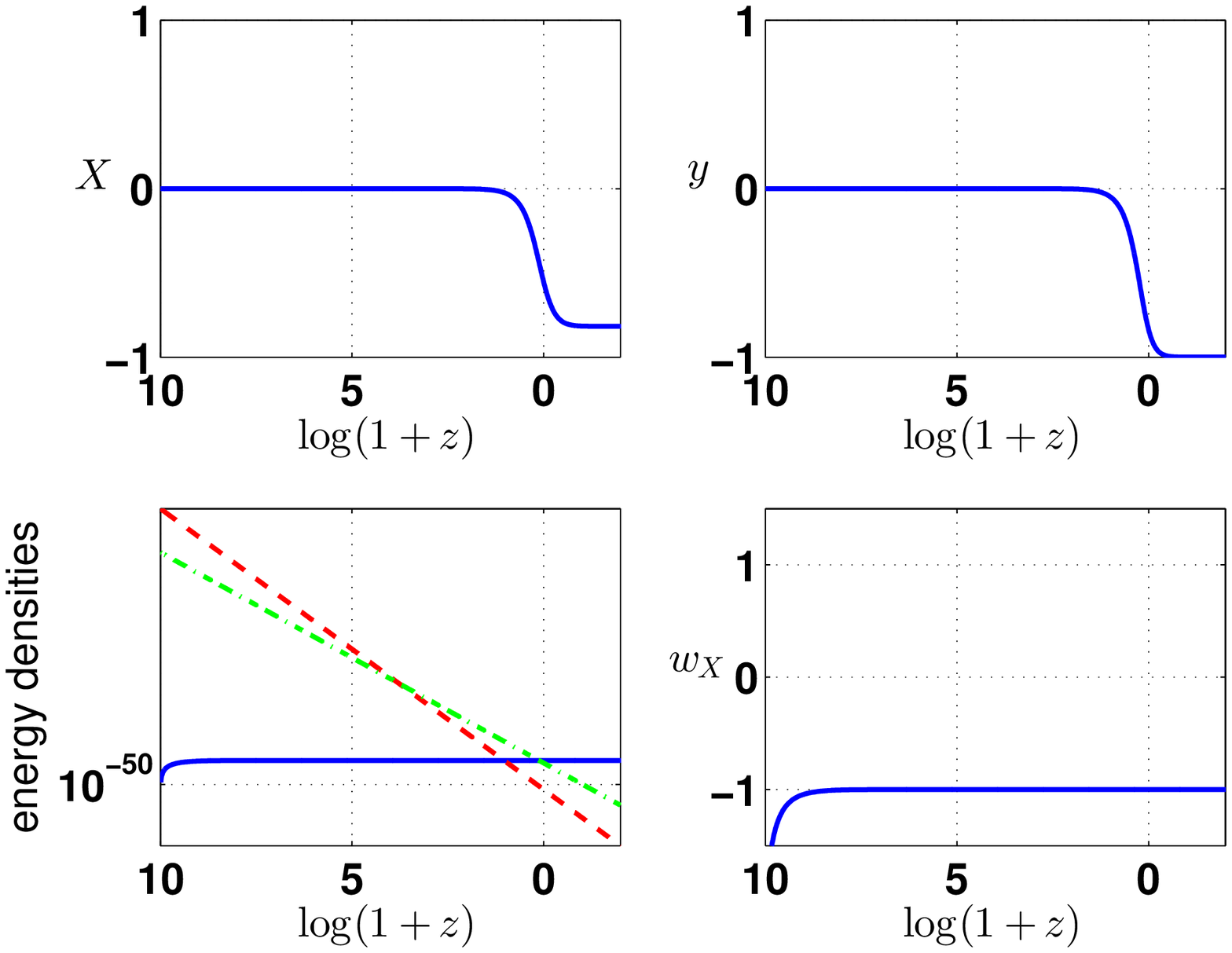}
\caption{\label{fig:case2a} Dynamics for a model without potential but with a linear coupling, $f = 1.62\times10^{-12}$ and $X_i=\dot{X}_i=0$. The different panels show the evolution of $X$, $y$; the energy density of the various components of the Universe; and the equation of state parameter $w_X$ with redshift. In the lower-left panel, the lines represent the energy density of radiation (dashed line), dust (dotted-dashed line) and $X$ (solid line). In this scenario, the Universe evolves from the fixed point $A$ to the fixed point $B$.}
\end{center}
\end{figure}

We now give a counter example and consider the case when $X_i=\dot{X}_i=0$ illustrated in Fig.~\ref{fig:case2a}. It can be observed that the Universe comes to be dominated by a three-form even though it emerged from a vanishing energy density. 

\begin{figure}[h]
\begin{center}
\includegraphics[width=0.6\columnwidth]{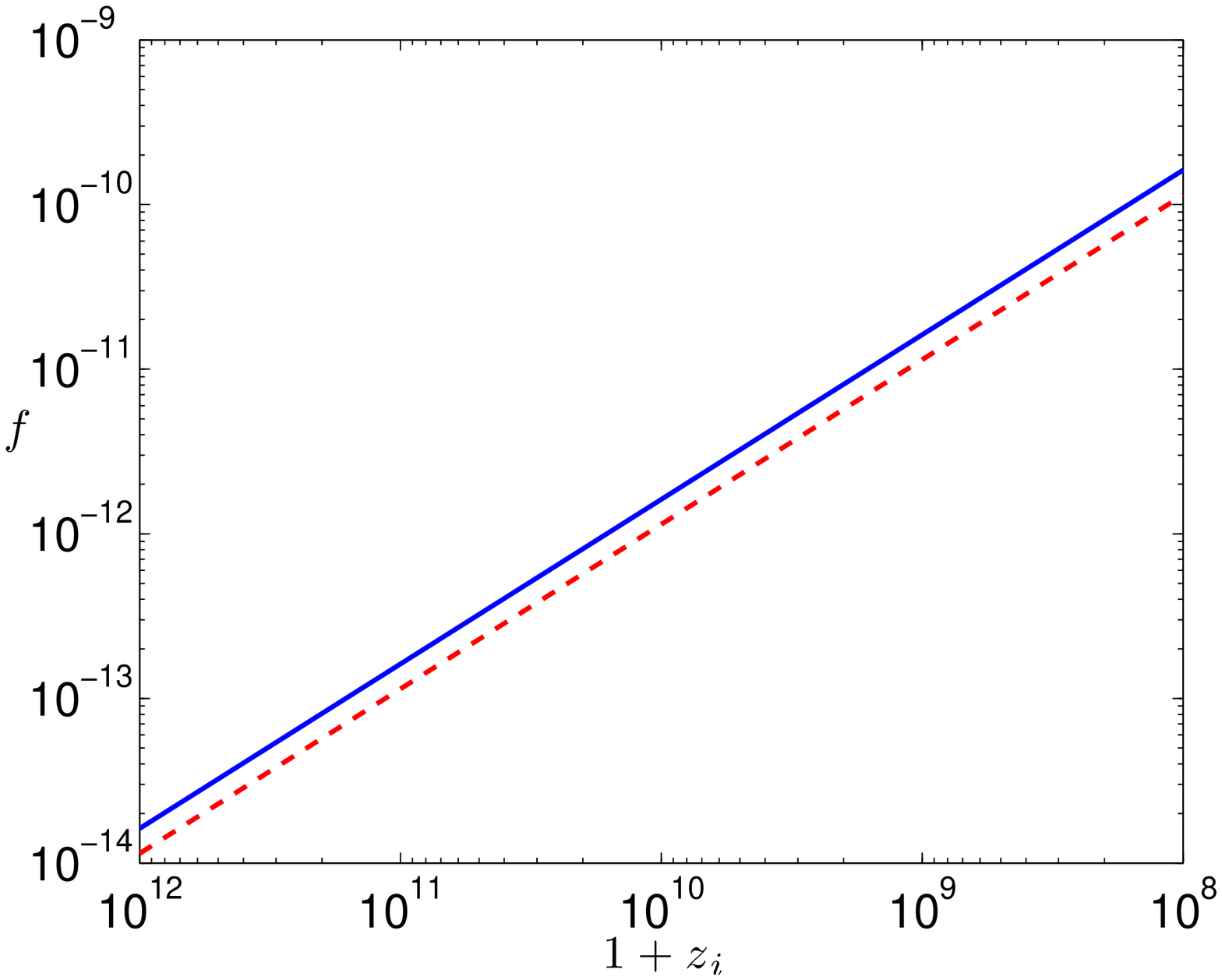}
\caption{\label{festimatefig} The solid line represents the value of $f$ required to obtain $\Omega_X = 0.7$ today and the dashed line represents the estimated value using Eq.~(\ref{festimateeq}).} 
\end{center}
\end{figure}

It is possible to make a quick estimate of the value of $f$ such that the Universe started to accelerate recently. Given that the dynamics of the Universe is dominated by a background perfect fluid with equation of state parameter $\gamma-1 = p_B/\rho_B$, the equation of motion for $X$ can be rewritten in terms of the dimensionless variable $y \equiv \kappa(X'+3X)/\sqrt{6}$ as
\be
\label{eomy}
y' -\frac{3}{2} \gamma y + \sqrt{\frac{3}{2}} \frac{\Omega_m}{\Omega_B} f e^{3(\gamma-1)N} = 0,
\ee
where $N = \ln(a/a_0)= -\ln(1+z_i)$, the prime means differentiation with respect to $N$,  $\gamma = 4/3$ for radiation, $\gamma = 1$ for dust, and we have neglected the contribution of the coupling in the energy density of dust; i.e., we have used $\rho_m = \rho_0 \Omega_m \exp(-3N)$. This equation of motion for $y$ has a simple solution 
\be
y = C e^{3\gamma N/2} + \frac{2}{\sqrt{6}(2-\gamma)} \frac{\Omega_m}{\Omega_B} f e^{3(\gamma-1) N},
\ee
where $C$ is a constant of integration which is set by the initial conditions.
We now take $y(N_i) = 0$ and match the solutions during matter domination and dust domination at $N_{eq} = \ln(\Omega_r/\Omega_m)$. In addition, we consider that $\rho_X$ becomes dominant at $N_a = \ln(\Omega_m/\Omega_X)^{1/3}$, as the equation of state parameter for $X$ is $w_X \approx -1$ at late times. Putting all this together, we can find a simple estimate for the value of $f$ such that the 3-form field is accelerating the Universe today:
\be
\label{festimateeq}
f^2 \approx \frac{\Omega_r \Omega_X}{3 \Omega_m^2 (1+z_i)^2}.
\ee
In Fig.~\ref{festimatefig} we show how this estimate compares with the value obtained by numerically solving the equation of motion and demanding $\Omega_X = 0.7$ today.
The degree of fine tuning is comparable to scalar field models of dark 
energy where the transition from a matter dominated to an accelerated 
Universe is controlled by the parameters of the scalar potential.

As the coupling parameter $f$ is very small, the exponential form of the coupling is not crucial for our conclusions, but the dynamics would be similar for any coupling that can be approximated by $m^2 \approx m_0 + f_0\kappa A + \dots$. 

We point out that the equation of state parameter shown in the  example of Fig.~\ref{fig:case2a} approaches $-1$ from below which might indicate the presence of a ghost in this particular case. For this reason, in what follows, we well avoid initial conditions leading to an equation of state parameter smaller than $-1$.

\subsection{Model with quadratic potential}

\begin{figure}[h]
\begin{center}
\includegraphics[width=0.8\columnwidth]{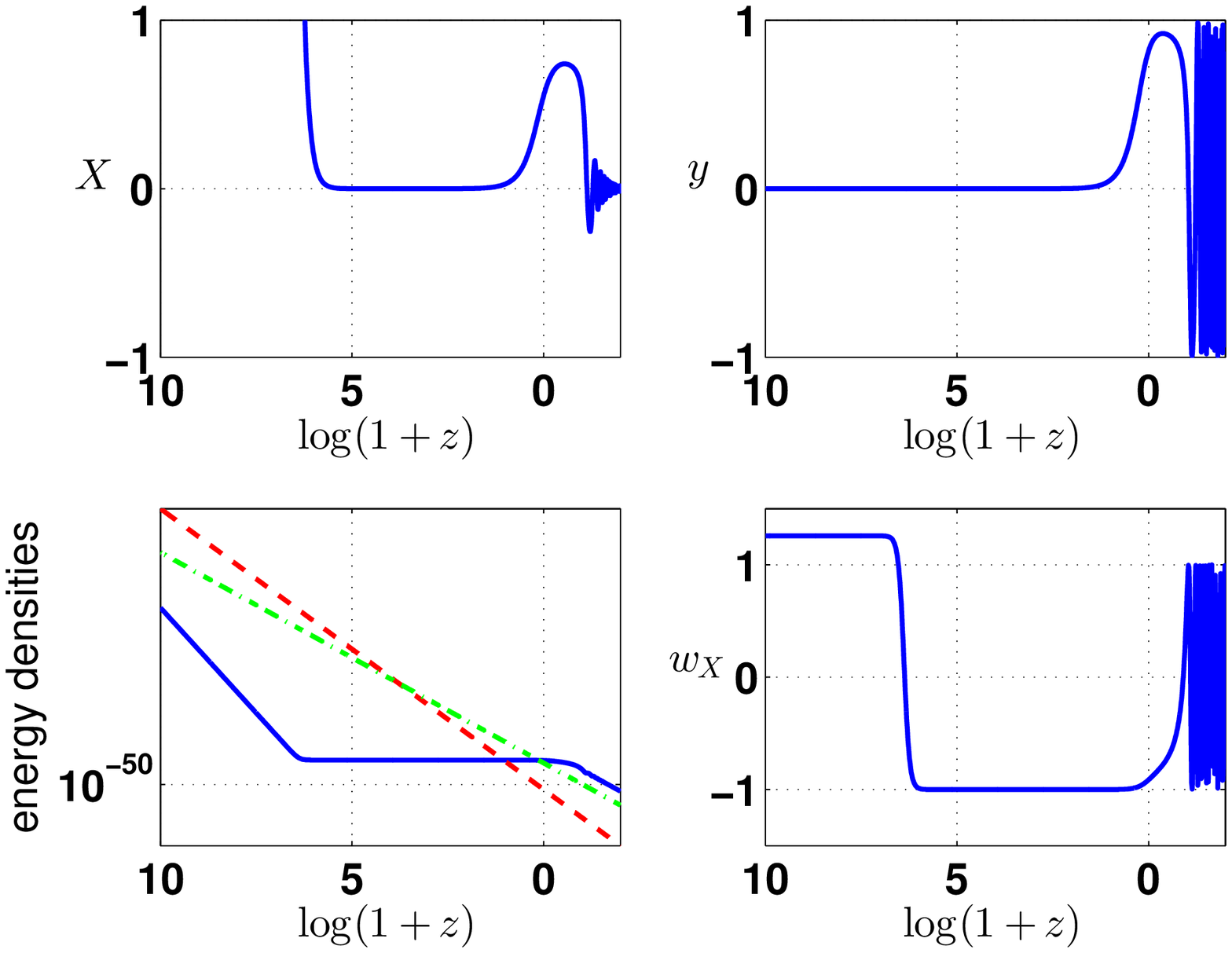}
\caption{\label{fig:case4} Dynamics for a model with a quadratic potential, $V = V_0 (\kappa X)^2$ with $V_0 = 4 \times 10^{-48}$GeV$^4$, and a linear coupling, $f = 1.725\times10^{-20}$. We used $\kappa X_i = 2 \times 10^{11}$ and $y_i = 1.2 \times 10^{-18}$. The different panels show the evolution of $X$, $y$; the energy density of the various components of the Universe; and the equation of state parameter $w_X$ with redshift. In the lower-left panel the lines represent the energy density of radiation (dashed line), dust (dotted-dashed line) and $X$ (solid line).}
\end{center}
\end{figure}

Considering now a quadratic potential for the three form such that $V = V_0 X^2$, the late dynamics of the Universe can be fairly different. The presence of the coupling between the three-form and dark matter displaces the field $X$ from the minimum toward $X \approx \pm \sqrt{2/3}$, driving the Universe to an accelerated expansion. When the energy density of the background decays, the field oscillates in the minimum of the potential, and the evolution of the Universe once again decelerates. The evolutions of $X$, $y$; the energy density of the various components of the Universe and the equation of state parameter $w_X$ with redshift are illustrated in Fig.~\ref{fig:case4}.
As in the example of Fig.~\ref{fig:case2}, the initial value of $w_X$ is larger than unity. This is a feature not present in the case when the coupling is not present \cite{Koivisto:2009ew}.

\begin{figure}[h]
\begin{center}
\includegraphics[width=0.6\columnwidth]{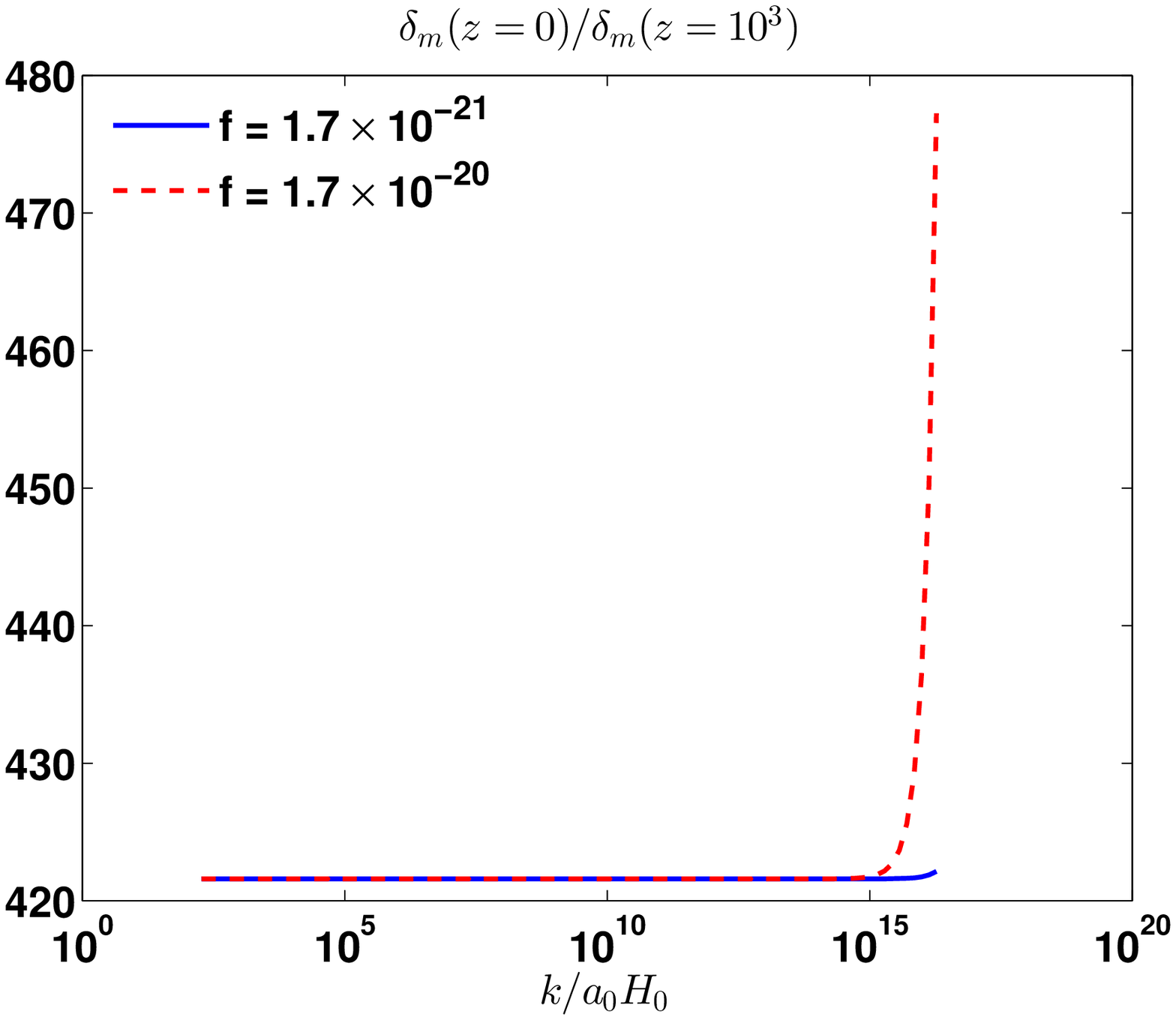}
\caption{\label{fig:growthk} Ratio of the linear density contrast $\delta_m(z=0)/\delta_m(z = 10^3)$ for scales that at $z = 10^3$ are $k = (10-10^{15}) aH$ and for two values of the coupling $f$.}
\end{center}
\end{figure}

When the potential is included, $\Sigma$ in Eq.~(\ref{deff}) is well defined, and we can use Eq. (\ref{deltaeq}) to compute the evolution of the matter density contrast. The result is illustrated, for two values of the coupling $f$, in Fig.~\ref{fig:growthk} as the ratio $\delta_m(z=0)/\delta_m(z = 10^3)$ for scales that at $z = 10^3$ are $k = (10-10^{15}) aH$.  The figure shows that there is hardly any difference in $\delta_m$ for the various scales, however, below a certain scale; i.e., for large enough $k/a$, the growth of perturbations increases abruptly. To understand this, we point out that $\Sigma$ takes negative values, and consequently, $c_{\rm eff}^2 < 0$, which means that there is an instability in the evolution of the linear density contrast when $(k/a)^2 > \kappa_{\rm eff}^2 \rho_m \delta_m/2 c_{\rm eff}^2$.    
 The conclusion is that the dimensionless coupling paramater $f$ is
constrained to be very small. Otherwise, as apparent in  Fig.~\ref{fig:growthk} , there is an explosive growth of the dark matter overdensity due to the three-form mediated fifth force for large enough $k$. If the
coupling is small enough, the critical wavelengths $k$ can correspond to
scales not probed by cosmology and outside the scope of linear perturbation
theory.

\begin{figure}[h]
\begin{center}
\includegraphics[width=0.6\columnwidth]{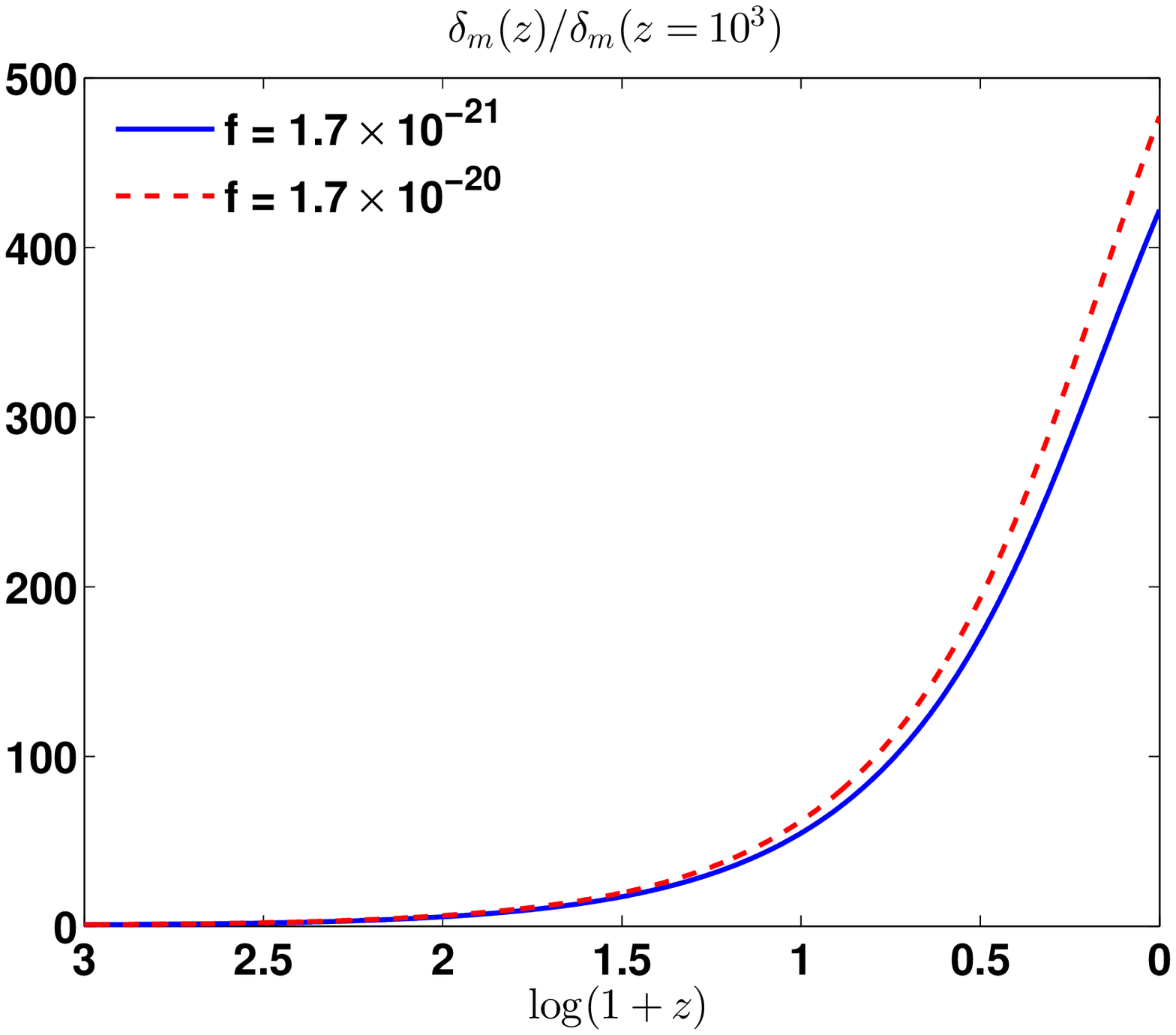}
\caption{\label{fig:growtht} Redshift dependence of $\delta_m$ for scales that at redshift $z = 10^3$ are $k =  10^{15} aH$ for two values of the coupling $f$.}
\end{center}
\end{figure}

In Fig.~\ref{fig:growtht} we show the evolution $\delta_m(z)$ with redshift for the same two values of $f$ used in Fig.~\ref{fig:growthk} and for a given scale $k$. It shows  that the evolution of the density contrast is faster for a larger value of $f$.

\section{Conclusions}
\label{concs}

We considered coupled three-form cosmologies. In particular, the cosmological fluid was modeled as a gas of point particles for which the mass depends upon the three-form. We presented the field equations and derived the cosmological equations of motion for both the background and perturbations. 

For the background, we performed a phase space analysis generalizing the previous results, finding a new class of fixed points and modified stability conditions due to the presence of the coupling. We also presented an example model where the massless three-form, initially vanishing, acquires a small, constant energy density as the coupling sets it evolving and until the field freezes in the fixed point $B$. The energy density would match the observed cosmological constant provided only that the coupling strength was sufficiently small (around ten-twenty orders of magnitude below the Planck scale depending on when inflation ended). 

In this work we ignored any loop effects arising from the
coupling of dark matter to the three-form field. These could in fact spoil
the three-form potential. We note, however, that qualitatively our
results apply regardless of the slope of the potential, and moreover, we
can obtain an accelerating Universe even in the absence of a potential.

For linear perturbations, we analyzed the features at the Newtonian limit and found the evolution of structure to be modified by an effective friction term, a time-varying effective gravitational coupling between the particles and a sound speed -like effect in the presence of which dark matter is not precisely cold. Since the perturbations are so sensitive to coupling, we might see subtle effects even with a very small coupling constant as in the one specific model considered here. In particular, the appearance of an effective nonzero sound speed introduces scale-dependence to structure formation. The coupling effects will be enhanced at small scales, which in the best case could help to alleviate such issues in the small-scale $\Lambda$CDM structure formation as the missing satellite problem.

The study of large-scale structure constraints in specific models, the workings of possible screening mechanisms and disformal couplings \cite{Koivisto:2012za} remain to be explored in the context of interacting three-form cosmology.

\section*{Acknowledgements}
We thank Anupam Mazumdar and Danielle Wills for useful suggestions. TK is supported by the Norwegian Research Council. NJN is supported by a Ci\^encia 2008 research contract funded by FCT and  through project No. PEst-OE/FIS/UI2751/2011 and  Grant No. CERN/FP/123615/2011.

\bibliography{C3f}

\end{document}